\documentclass[aps,prl,reprint,superscriptaddress]{revtex4-2}

\usepackage{graphicx}
\usepackage{dcolumn}
\usepackage{bm}

\begin{document}

\preprint{v2}


\title{Current-induced magnon trapping in spin torque oscillation}



\author{Takahiko Makiuchi}
\email{makiuchi@g.ecc.u-tokyo.ac.jp}
\affiliation{Quantum-Phase Electronics Center, the University of Tokyo, Tokyo 113-8656, Japan.}
\author{Naoki Kanazawa}
\affiliation{IBM Quantum, IBM Research -- Tokyo, Tokyo 103-8510, Japan.}
\author{Eiji Saitoh}
\affiliation{Quantum-Phase Electronics Center, the University of Tokyo, Tokyo 113-8656, Japan.}
\affiliation{Department of Applied Physics, the University of Tokyo, Tokyo 113-8656, Japan.}
\affiliation{Institute for AI and Beyond, the University of Tokyo, Tokyo 113-8656, Japan.}
\affiliation{RIKEN Center for Emergent Matter Science (CEMS), Wako 0351–0198, Japan.}
\affiliation{Advanced Institute for Materials Research, Tohoku University, Sendai 980-8577, Japan.}
\affiliation{Advanced Science Research Center, Japan Atomic Energy Agency, Tokai 319-1195, Japan.}


\date{\today}

\begin{abstract}
 Spin torque nano-oscillators realized by magnetization dynamics trapped in a current-induced potential are reported.
 We fabricated Ni$_{81}$Fe$_{19}$/Pt nanostructures and measured current-induced microwave emission from the structures.
 The result shows an increase in the magnitude and spectral narrowing of the microwave emission.
 We demonstrate that the current-induced magnetic field suppresses magnon radiation loss and significantly reduces the linewidth and the threshold current required for the spin torque oscillation.
 \end{abstract}

\keywords{Spin Hall nano-oscillator, spin torque nano-oscillator, microwave generation, low temperature, nonlinear effect, magnetization dynamics, spintronics}

\maketitle

Spin torque oscillators have gained significant interest in their ability to convert direct current into microwave \cite{slonczewski1996current, berger1996emission, kiselev2003microwave, slavin2009nonlinear}.
This conversion occurs when the spin-transfer torque, which is proportional to the intensity of the direct current, surpasses a damping and initiates magnetic oscillation.
One of the primary contributors to the damping is the diffusion of magnons beyond the excitation region \cite{demidov2011control}. 
While one approach to address this issue is to fabricate small magnetic structures, this method often introduces defects during processing, which can serve as additional sources of damping.

In this study, we demonstrate that the direct current not only induces the magnetic oscillation but also stabilizes it through the Oersted field (induction magnetic field).
To clarify the effect of the Oersted field, we utilize spin torque nano-oscillators (STNOs) comprising a ferromagnetic permalloy (Ni$_{81}$Fe$_{19}$ or NiFe) and heavy-metal platinum (Pt) bilayer \cite{demidov2012magnetic, liu2013spectral, demidov2014synchronization, ulrichs2014micromagnetic, hamadeh2014full, chen2019dynamical}.
We change the profile of current-induced Oersted field by employing two types of STNOs with reversed layer orders: NiFe/Pt (Ni$_{81}$Fe$_{19}$ on Pt) and Pt/NiFe.
We show that only the NiFe/Pt STNO exhibits a trapping Oersted field for spin waves (magnons).
Spectroscopic experiments reveal that the NiFe/Pt configuration yields a larger microwave intensity, narrower linewidth, and smaller threshold current, attributed to the current-induced trapping effect on spin waves.
The NiFe/Pt configuration exhibits a localized magnetic oscillation, whereas the Pt/NiFe configuration, with an anti-trapping Oersted field, shows extended and unstable oscillation profiles.

\begin{figure}
 \includegraphics[width=86mm]{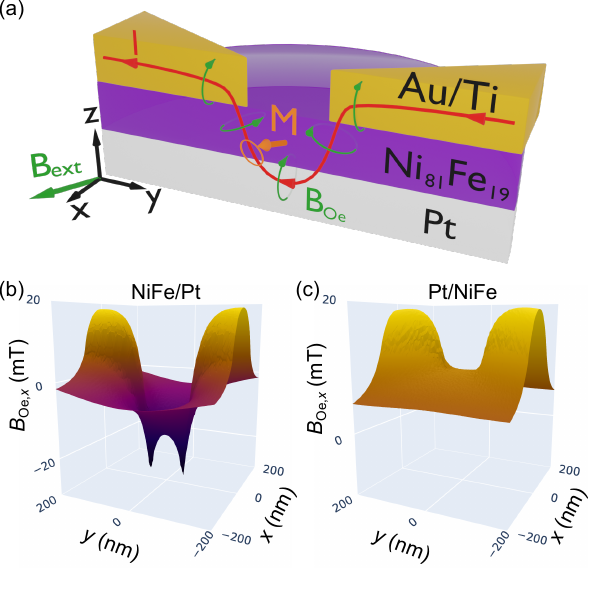}%
 \caption{(a) Cross-sectional illustration of a Ni$_{81}$Fe$_{19}$/Pt STNO. (b,c) Calculated Oersted field in the Ni$_{81}$Fe$_{19}$ layer for the NiFe/Pt and Pt/NiFe STNOs, respectively. The input direct current is $I= -20$ mA.}
 \label{fig:COMSOL}
\end{figure}

First, we calculate the spatial profile of the Oersted field for three-dimensional STNO models by using the finite element method (COMSOL Multiphysics).
The device consists of an extended bilayer disk of Ni$_{81}$Fe$_{19}$(5 nm)/Pt(5 nm) on which triangular Au(80 nm)/Ti(10 nm) electrodes are formed, as shown in Fig. \ref{fig:COMSOL}(a).
The values in the parentheses are the thicknesses.
The diameter of the bilayer disk is 4 µm.
Each Au/Ti electrode has a round tip of 50 nm in radius, and the pair of electrodes is placed with a 100 nm gap.
A direct current of $I=-20$ mA (current from $+y$) is supplied to the electrodes.
The electrical conductance values of $\sigma_\mathrm{Au} > \sigma_\mathrm{Pt} > \sigma_\mathrm{Ti} > \sigma_\mathrm{Ni_{81}Fe_{19}}$ (45.6, 8.9, 2.6, 1.74 MS/m, respectively) are used.
A large portion of the direct current passes through the Au/Ti electrodes and goes into the Pt layer near the nano-gap region.
The simulation yields the Oersted field and charge current density.

Figures \ref{fig:COMSOL}(b) and \ref{fig:COMSOL}(c) illustrate the Oersted fields in the NiFe layer of the NiFe/Pt and Pt/NiFe STNO, respectively.
In the NiFe/Pt device, a local dip appears in the nano-gap region where the current penetrates into the underlying Pt layer [Fig. \ref{fig:COMSOL}(a)].
The dip of the Oersted field serves as a trapping potential for spin waves \cite{demidov2014nanoconstriction, verba2019interplay, noack2021evolution}.
In contrast, the Pt/NiFe STNO exhibits a saddle-like field profile in the proximity of the nano-gap.

The current density profiles ($j_x,\ j_y$) in the Pt layer are similar in both NiFe/Pt and Pt/NiFe devices.
The current density reaches a minimum of approximately $j_y \approx -6\times 10^{12}\ \mathrm{A/m^2}$ at the center of the nano-gap in both cases.
Consequently, the same value of current $I$ yields a similar amount of spin-transfer torque, irrespective of the layer order.
Therefore, any observed differences in experiments and simulations are likely to stem from variations in the Oersted field.

\begin{figure}
 \includegraphics[width=86mm]{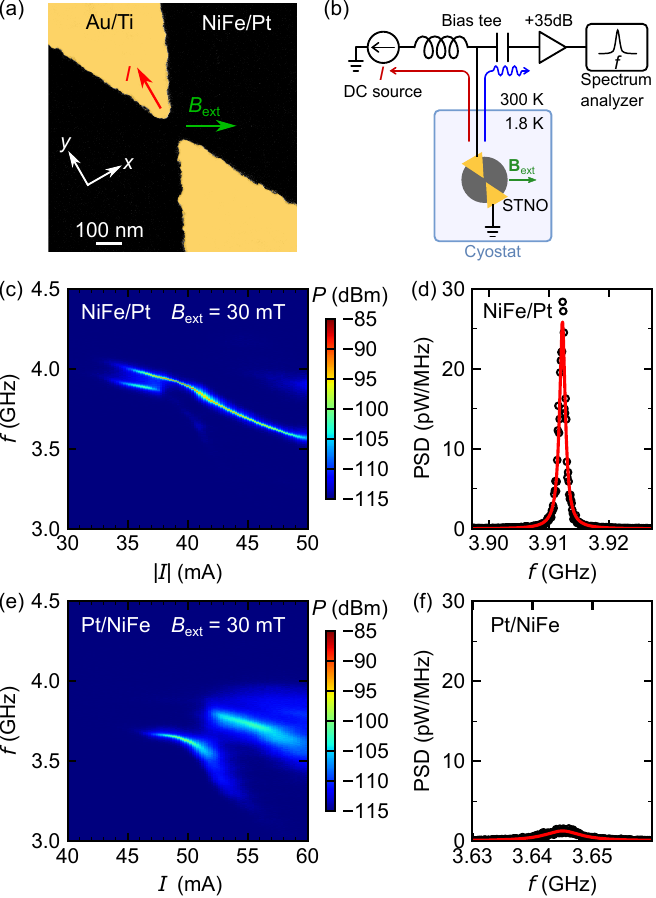}
 \caption{(a) Scanning electron microscope image of the NiFe/Pt STNO. The gap is 110 nm. The arrows indicate positive directions of $I$ and $B_\mathrm{ext}$. (b) Measurement circuit. (c) Microwave signal from the NiFe/Pt STNO. (d) Highest $Q$ power spectral density (PSD) from the NiFe/Pt STNO. The red curve is a Lorenzian fitting. $B_\mathrm{ext}=30$ mT; $I=-$ 39.2 mA; $Q=6900$. (e) Microwave signal from the Pt/NiFe STNO. (f) Highest $Q$ PSD from the Pt/NiFe STNO. $B_\mathrm{ext}=30$ mT; $I=$ 50.8 mA; $Q=990$.}
 \label{fig:expt}
\end{figure}

To investigate the impact of the Oersted field on magnetic oscillation, we fabricate two types of STNO and conducted spectroscopic measurements.
We prepare sputtered Ni$_{81}$Fe$_{19}$(5 nm)/Pt(5 nm) and Pt(5 nm)/Ni$_{81}$Fe$_{19}$(5 nm) films on separate sapphire substrates.
Each film is shaped into a 4-µm-diameter disk with electron beam lithography and Ar ion milling techniques.
We make a pair of nano-gapped Au(80 nm)/Ti(10 nm) electrodes on each disk by using an electron beam lithography and sputtering.
Figure \ref{fig:expt}(a) is a scanning electron microscope image of the NiFe/Pt STNO.
Both NiFe/Pt and Pt/NiFe STNOs have the same Au/Ti electrode gap of 110 nm.
The electrodes are wire-bonded to a waveguide that introduces the direct current and receives the microwave generated by the STNO.

The STNO is placed inside a cryostat with a superconducting magnet.
The angle between the current $I$ and the external field $B_\mathrm{ext}$ directions is set to 120 degrees as shown in Fig. \ref{fig:expt}(a), which is the optimum condition for the microwave generation in terms of the spin polarization direction and anisotropic magnetoresistance \cite{liu2013spectral, chen2019dynamical}.
The cryogenic environment helps to decrease the magnetic damping.
The ambient temperature is kept at 1.8 K while the STNO is heated above 10 K during the current injection, inferred from the device resistance.
We inject the direct current and retrieved the microwave signal from the STNO with a bias tee.
The microwave is amplified and measured with a spectrum analyzer, as shown in \ref{fig:expt}(b).
The gain and loss in the microwave components are subtracted from the presented data.

Figures \ref{fig:expt}(c) and \ref{fig:expt}(e) depict the microwave power spectrum as a function of the direct current for the NiFe/Pt and Pt/NiFe STNOs, respectively.
The direction of the external field $\mathbf{B}_\mathrm{ext}$ is fixed, and the sign of the direct current $I$ is reversed for the two STNOs to observe microwave signals.
The signs of $I$ align consistently with the anti-damping spin-transfer torque direction for the different layer orders.
The threshold currents for the onset of the microwave signal are remarkably different.
The signal appears above $|I_\mathrm{th}|=$ 33 mA in Fig. \ref{fig:expt}(c), whereas it requires a notably larger current of $I_\mathrm{th}=$ 46 mA in Fig. \ref{fig:expt}(e).
Two modes coexist at 33 mA $<|I|<$ 38 mA in Fig. \ref{fig:expt}(c).
The higher frequency mode survives above 38 mA and is entrained to the lower frequency.
The different frequency modes are attributed to modes with different spatial profiles of the oscillating magnetic moments.
The lower and higher frequency modes can be assigned as bullet and quasi-linear modes, respectively \cite{slavin2005spin, ulrichs2014micromagnetic, chen2019dynamical, divinskiy2019controllable}.
The bullet mode is a localized mode that, in principle, does not propagate spin waves \cite{slavin2005spin}.
The quasi-linear mode is not as strongly detuned as the bullet mode from the linear magnon dispersion, allowing it to radiate spin waves.
In contrast, the two modes in Fig. \ref{fig:expt}(e) are broader and do not coexist at the same $I$.
The coexistence of the two modes [\ref{fig:expt}(c)] indicates that the mode coupling is strong enough to support energy exchange between them, whereas their separation [\ref{fig:expt}(e)] indicates that only the mode with a smaller threshold current is realized due to weak coupling resulting from spatial separation.

All the modes in Figs. \ref{fig:expt}(c) and \ref{fig:expt}(e) have lower frequency than the linear mode frequency expressed by the Kittel formula $\omega= \gamma \sqrt{B_\mathrm{ext} (B_\mathrm{ext} + \mu_0 M_\mathrm{s})} = 2\pi \times 4.3$ GHz.
These redshifts are consistent with the nonlinear frequency shift \cite{slavin2005spin, liu2013spectral} and/or the reduction of the local resonant frequency by the Oersted field.
Here, $\gamma$ is the gyromagnetic ratio, $\mu_0$ is the permeability, and $M_\mathrm{s}$ is the saturation magnetization.
We obtained $\mu_0M_\mathrm{s} = 760$ mT from a spin-torque ferromagnetic resonance measurement on a similar STNO device \cite{tulapurkar2005spin, sankey2006spin, liu2011spin}.

The power spectral density (PSD) with the highest quality (Q) factor for each device is plotted in Figs. \ref{fig:expt}(d) and \ref{fig:expt}(f).
There is an order of magnitude difference between the maximal PSDs (26 pW/MHz for the NiFe/Pt device and 1.3 pW/MHz for the Pt/NiFe device) and their corresponding Q factors ($Q= 6900$ for the NiFe/Pt device and 990 for the Pt/NiFe device).
These essential improvements in the NiFe/Pt device can be attributed to the effect of the Oersted field trapping, by which the localized mode cannot diffuse outside the excited region.

We conduct micromagnetic simulations using MuMax3 \cite{vansteenkiste2014design} with geometry closely resembling the experimental setup.
The Ni$_{81}$Fe$_{19}$ layer is a 5-nm-thick disk of 4 µm in diameter.
The entire system is divided into 1024 $\times$ 1024 $\times$ 1 rectangular cells, forming a 4000 nm $\times$ 4000 nm $\times$ 5 nm cuboid space.
The applied in-plane external field $\mathbf{B}_\mathrm{ext}$ is tilted at an angle of 120 degrees relative to the current direction and $|\mathbf{B}_\mathrm{ext}|=30$ mT, as in the experiment.
The effect of the Pt layer is taken into consideration to provide spatially-varying spin current and Oersted field $\mathbf{B}_\mathrm{Oe}$ into the Ni$_{81}$Fe$_{19}$ layer.
The Slonczewski anti-damping spin-transfer torque acting on the Ni$_{81}$Fe$_{19}$ magnetization is described by \cite{vansteenkiste2014design}
\begin{equation}
 \bm{\tau} \approx \frac{\theta_\mathrm{SH} j_\mathrm{c} \gamma \hbar P}{2M_\mathrm{s}ed} \mathbf{m}\times (\mathbf{m}_\mathrm{P}\times\mathbf{m}),
\end{equation}
where $\theta_\mathrm{SH}=0.07$ is the spin Hall angle of Pt, ($j_x,\ j_y$) is the current density profile in the Pt layer with $j_\mathrm{c}=\sqrt{j_x^2 +j_y^2}$, $\hbar$ is the Dirac constant, $P=1$ is the spin polarization, $M_\mathrm{s}=597$ kA/m 
is the saturation magnetization, $e$ is the elementary charge, $d=5$ nm is the Ni$_{81}$Fe$_{19}$ thickness, $\mathbf{m}$ is the reduced magnetization, and $\mathbf{m}_\mathrm{P} = \pm(-j_y/j_\mathrm{c}, j_x/j_\mathrm{c}, 0)$ is the spin polarization direction.
The sign in $\mathbf{m}_\mathrm{P}$ depends on the layer order ($-$ for NiFe/Pt and $+$ for Pt/NiFe).
Spatial profiles of $j_\mathrm{c}$, $\mathbf{m}_\mathrm{P}$ and $\mathbf{B}_\mathrm{Oe}$, calculated with COMSOL, are incorporated in the MuMax3 simulation.
$j_\mathrm{c}$ and $\mathbf{B}_\mathrm{Oe}$ for a given $I$ are obtained by using a proportionality relation: $j_\mathrm{c}$, $B_{\mathrm{Oe},x}$, $B_{\mathrm{Oe},y}$, $B_{\mathrm{Oe},z} \propto I$.
Other parameters utilized are the exchange constant $A_\mathrm{ex}=13$ pJ/m and Gilbert damping coefficient $\alpha = 0.02$.
We set $\alpha$ to 1 at the Ni$_{81}$Fe$_{19}$ disk edge to account for damping affected by the edge roughness.
The system temperature is 0 K for simplicity.
We confirm that introducing a finite temperature thermal fluctuation \cite{vansteenkiste2014design} did not qualitatively change the results.

\begin{figure}
 \includegraphics[width=86mm]{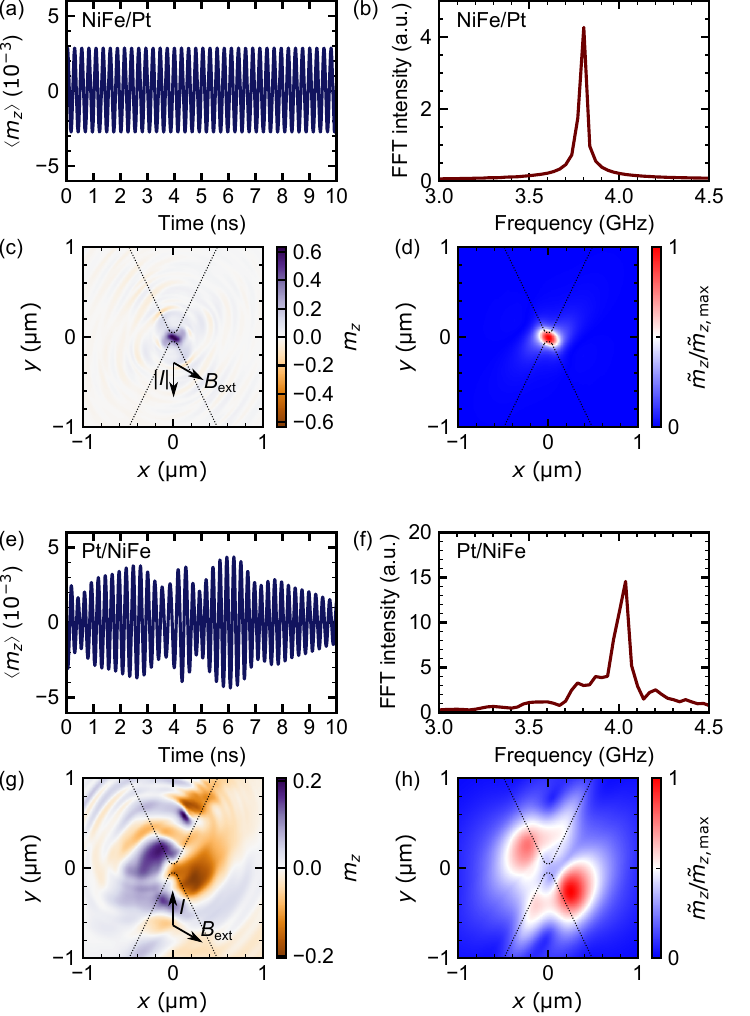}%
 \caption{Micromagnetic simulation for the (a--d) NiFe/Pt and (e--h) Pt/NiFe devices. (a,e) Temporal dependence of the spatially-averaged out-of-plane magnetization $\langle m_z\rangle$. (b,f) Fourier spectrum of $\langle m_z\rangle$ over a 2.5 µm $\times$ 2.5 µm square area with a time window of 30 ns. (c,g) Snapshot of $m_z(x,y)$. Dashed curves represent the edge of electrodes. (d,h) FFT intensity of $m_z(x,y)$ with a frequency window of 3.6--4.0 (NiFe/Pt) and 3.8--4.2 GHz (Pt/NiFe).}
 \label{fig:micromag}
\end{figure}

Figure \ref{fig:micromag} shows the results of the micromagnetic simulations.
The applied direct current intensity is set a few milliamperes above the threshold current: $I=-18$ mA for the NiFe/Pt and 44 mA for the Pt/NiFe case, respectively.
The greater threshold current for Pt/NiFe indicates an additional dissipation channel.

Figures \ref{fig:micromag}(a) and \ref{fig:micromag}(e) show the time evolution of a spatial average of the normalized out-of-plane magnetization component $m_z(x,y,t)$.
Both cases show a similar period of about 0.25 ns ($\sim 4$ GHz).
However, the NiFe/Pt device manifests stationary oscillation, whereas the Pt/NiFe device exhibits a non-monotonic envelope.

The fast Fourier transform (FFT) spectrum in Fig. \ref{fig:micromag}(b) exhibits a distinct peak at 3.8 GHz for the NiFe/Pt configuration.
Conversely, the FFT spectrum in Fig. \ref{fig:micromag}(f) corresponding to the Pt/NiFe device displays a broad peak around 4.0 GHz, indicating poor coherence.
Both the simulations [Fig. \ref{fig:micromag}(b) \& \ref{fig:micromag}(f)] and experiments [Fig. \ref{fig:expt}(c) \& \ref{fig:expt}(e), at relatively large currents] yield the low (high) frequency mode for the NiFe/Pt (Pt/NiFe).
We do not observe a second peak with $|\mathbf{B}_\mathrm{ext}|=$ 30 mT in the simulations.
This indicates that the supplied spin-transfer torque excites only the main mode and the energy does not transfer from the main to the second mode.

The snapshot of $m_z(x,y,t)$ in Fig. \ref{fig:micromag}(c) represents the auto-oscillation near the nano-gap in the NiFe/Pt device.
The mode is confined and has a large amplitude within the nano-gap.
This mode is seen in the FFT intensity image in Fig. \ref{fig:micromag}(d).

In contrast, the mode in the Pt/NiFe case has amplitudes and wavefronts around the nano-gap in Fig. \ref{fig:micromag}(g).
The waves move outward with a velocity of the order of 1 µm/ns.
This radiative feature is attributed to the reason for the large threshold current.
The spatial profile of the FFT intensity, shown in Fig. \ref{fig:micromag}(h), has a small amplitude at the nano-gap center and a relatively large amplitude on one side of each electrode, suggesting that a mode generated at the current concentrated center is delocalized.

The difference in the threshold current is attributed to the effect of the self-induced trapping field;
the simulations of the Pt/NiFe and NiFe/Pt STNOs differ in their Oersted fields, but the current densities are similar for a given current magnitude $I$.
The confined spin-torque oscillation in the NiFe/Pt is a manifestation of the trapping effect, whereas the extended oscillation in the Pt/NiFe case is due to the anti-trapping by the saddle-like Oersted field profile.

The spatial variation of the Oersted field changes the local resonant frequency
\begin{equation}
    \omega(k) = \sqrt{(\omega_B + Dk^2)(\omega_B + Dk^2 + \omega_M)},
\end{equation}
where $\omega_B = \gamma |\mathbf{B}_\mathrm{ext}+\mathbf{B}_\mathrm{Oe}|$, $\omega_M = \gamma \mu_0 M_\mathrm{s}$, $D=2\gamma A_\mathrm{ex}/M_\mathrm{s}$ is the exchange stiffness, and $k$ is the wavenumber.
In the NiFe/Pt device, the resonant frequency exhibits a local minimum at the nano-gap region due to the specific Oersted field profile depicted in Fig. \ref{fig:COMSOL}(b).
This trapping field confines the magnetization dynamics within the nano-gap, preventing their escape  \cite{demidov2014nanoconstriction, verba2019interplay, noack2021evolution}.
In contrast, the Pt/NiFe device demonstrates a saddle point in the Oersted field, as seen in Fig. \ref{fig:COMSOL}(c), enabling the magnetization dynamics to attain finite group velocities and escape from the nano-gap region.

\begin{figure}
 \includegraphics[width=60mm]{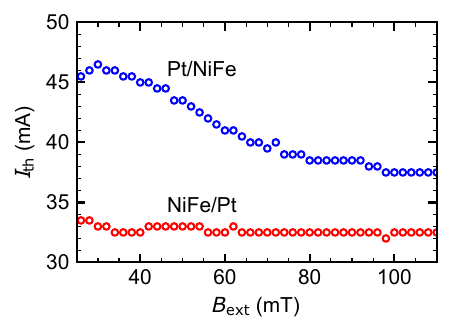}%
 \caption{The external field dependence of the threshold current for the microwave generation. $I_\mathrm{th}$ is defined as the minimum current required to achieve a microwave signal approximately 3 dB above the noise floor.}
 \label{fig:threshold}
\end{figure}

Figure \ref{fig:threshold} shows the threshold current $I_\mathrm{th}$ for the onset of spin-torque oscillation as a function of $B_\mathrm{ext}$, derived from the experimental data.
In the Pt/NiFe STNO configuration, $I_\mathrm{th}$ consistently surpasses that of the NiFe/Pt, particularly noticeable at lower external fields $B_\mathrm{ext}$.
The significant difference in $I_\mathrm{th}$ at small $B_\mathrm{ext}$ is attributed to the substantial contribution of the Oersted field to the effective magnetic field ($\mathbf{B}_\mathrm{ext} + \mathbf{B}_\mathrm{Oe}$).
As $B_\mathrm{ext}$ increases, $I_\mathrm{th}$ for Pt/NiFe reaches a saturation point, ceasing to decrease.
The finite gap between the threshold currents across the entire range of external fields signifies that selecting the appropriate order of magnetic and heavy metal layers contributes to stabilizing spin-torque oscillation in numerous scenarios.

The spatial profile from the simulation [Figs. \ref{fig:micromag}(d) \& \ref{fig:micromag}(h)] and the low/high frequency relations from both the experiment and simulation [Figs. \ref{fig:expt}(c), \ref{fig:expt}(e) and \ref{fig:micromag}(b) \& \ref{fig:micromag}(f)] indicate that the low-frequency localized bullet mode is facilitated by the trapping Oersted field.
In contrast, the anti-trapping Oersted field allows for the excitation of a high-frequency propagating mode around the region of current concentration.

Among various methods to stabilize spin torque oscillations, the Oersted field trap offers distinct advantages.
Unlike the dipolar field trap \cite{chia2012nanoscale, urazhdin2014nanomagnonic, zhang2021spin} or geometrical confinements such as nano-constriction \cite{demidov2014nanoconstriction, durrenfeld201720, divinskiy2017magnetic, dvornik2018origin} and nano-wire \cite{duan2014nanowire, verba2019interplay}, the current-induced magnon trapping by the Oersted field does not require additional boundaries or magnetic layers, which could increase the number of standing wave modes, stray fields and defects. 
Our study reveals that the Oersted field is not just an accessory \cite{demidov2011control, liu2013spectral, demidov2014nanoconstriction, ulrichs2014micromagnetic, durrenfeld201720, divinskiy2017magnetic, dvornik2018origin, chen2019dynamical} but a crucial factor for the stability of spin torque oscillation.
By adopting this approach, we may be able to attain coherent oscillations without the need for complex setups, making it an efficient and appealing choice for practical applications for low-temperature microwave generators.

In summary, our investigation delved into the impact of the current-induced Oersted field on the spin-torque oscillation.
In the experimental configuration wherein the ferromagnet Ni$_{81}$Fe$_{19}$ layer is above the heavy metal Pt layer, the microwave signal appears with a small threshold direct current, a large quality factor, and a large microwave power.
The characteristics observed in the time and frequency domain simulations validate the small threshold current and the localization of the spin-torque oscillation when influenced positively by the Oersted field.

\begin{acknowledgments}
 We thank K. Hoshi, T. Hioki, S. Daimon, M. Tokunari, K. Masuda, T. Itoko, and D. Nakano for fruitful discussions.
 This work was partially supported by IBM--UTokyo lab, JSPS KAKENHI (Nos. JP19H05600 and JP22H05114), JST CREST (Nos. JPMJCR20C1 and JPMJCR20T2) and JST ASPIRE (No. JPMJAP2316).
\end{acknowledgments}

\bibliographystyle{apsrev4-2.bst}

\end{document}